\documentclass[sigconf]{acmart}
\AtBeginDocument{%
  }

\usepackage{float}
\usepackage{comment}
\usepackage{listings}
\usepackage{multirow}
\usepackage{multicol}
\usepackage[]{mdframed}
\usepackage{tcolorbox}
\usepackage{booktabs}
\usepackage{colortbl}
\usepackage{xcolor}
\usepackage{subcaption}
\usepackage{balance}

\usepackage[font=small,labelfont=bf]{caption}
\definecolor{dkgreen}{rgb}{0,0.6,0}
\definecolor{gray}{rgb}{0.5,0.5,0.5}
\definecolor{mauve}{rgb}{0.58,0,0.82}
\definecolor{lightgrey}{RGB}{211, 211, 211}      %

\usepackage{xcolor}
\definecolor{headerBlue}{RGB}{51, 122, 183}      %
\definecolor{defnGray}{RGB}{70, 70, 70}          %
\definecolor{lightBg}{RGB}{247, 250, 252}        %
\definecolor{predicateColor}{RGB}{25, 103, 140}  %
\definecolor{varColor}{RGB}{70, 136, 71}         %
\definecolor{typeColor}{RGB}{156, 51, 151}       %

\definecolor{mainBlue}{RGB}{51, 122, 183}  %
\definecolor{lightBg}{RGB}{248, 251, 253}  %
\definecolor{borderColor}{RGB}{220, 226, 231}  
\definecolor{emphColor}{RGB}{70, 136, 71}  

\setcopyright{acmlicensed}
\copyrightyear{2025}
\acmYear{2025}
\acmDOI{XXXXXXX.XXXXXXX}

\acmConference[EASE 2025]{The 29th International Conference on Evaluation and Assessment in Software Engineering}
\acmISBN{978-1-4503-XXXX-X/18/06}

\begin{document}

\tcbset{
    myboxstyle/.style={
        colback=gray!10, %
        colframe=gray!70, %
        fonttitle=\bfseries, %
        coltitle=black, %
        boxrule=0.5mm, %
        arc=4mm, %
        left=0.2mm, %
        right=1mm, %
        top=0.5mm, %
        bottom=1mm, %
        width=8.4cm, %
        enlarge left by=0mm, %
        title={}, %
        before=\vspace{3mm}, %
        after=\vspace{3mm} %
    }
}

\title{Micro-Patterns in Solidity Code}

\author{Luca Ruschioni}
\email{luca.ruschioni@unicam.it}
\orcid{0000-0002-7330-4051}
\affiliation{%
  \institution{University of Camerino}
  \city{Camerino}
  \country{Italy}
}

\author{Robert Shuttleworth}
\email{robert.shuttleworth2@brunel.ac.uk}
\orcid{0009-0001-3930-4779}
\affiliation{%
  \institution{Brunel University of London}
  \city{London}
  \country{UK}
}

\author{Rumyana Neykova}
\email{rumyana.neykova@brunel.ac.uk}
\orcid{0000-0002-2755-7728}
\affiliation{%
  \institution{Brunel University of London}
  \city{London}
  \country{UK}
}

\author{Barbara Re}
\email{barbara.re@unicam.it}
\orcid{0000-0001-5374-2364}
\affiliation{%
  \institution{University of Camerino}
  \city{Camerino}
  \country{Italy}
}

\author{Giuseppe Destefanis}
\email{giuseppe.destefanis@brunel.ac.uk}
\orcid{0000-0003-3982-6355}
\affiliation{%
  \institution{Brunel University of London}
  \city{London}
  \country{UK}
}

\renewcommand{\shortauthors}{Ruschioni et al.}

\begin{abstract}
Solidity is the predominant programming language for blockchain-based smart contracts, and its characteristics pose significant challenges for code analysis and maintenance. Traditional software analysis approaches, while effective for conventional programming languages, often fail to address Solidity-specific features such as gas optimization and security constraints.
This paper introduces micro-patterns - recurring, small-scale design structures that capture key behavioral and structural peculiarities specific to a language - for Solidity language and demonstrates their value in understanding smart contract development practices. We identified 18 distinct micro-patterns organized in five categories (Security, Functional, Optimization, Interaction, and Feedback), detailing their characteristics to enable automated detection.
To validate this proposal, we analyzed a dataset of 23258 smart contracts from five popular blockchains (Ethereum, Polygon, Arbitrum, Fantom and Optimism). Our analysis reveals widespread adoption of micro-patterns, with 99\% of contracts implementing at least one pattern and an average of 2.76 patterns per contract. The \textit{Storage Saver} pattern showed the highest adoption (84.62\% mean coverage), while security patterns demonstrated platform-specific adoption rates. Statistical analysis revealed significant platform-specific differences in adoption, particularly in \textit{Borrower}, \textit{Implementer}, and \textit{Storage Saver} patterns.

\end{abstract}



\maketitle

\section{Introduction}\label{introduction}

Solidity is a statically-typed, object-oriented programming language designed for implementing smart contracts on blockchain platforms. Since its release in 2015, Solidity has become the primary language for blockchain development \cite{antonopoulos2018mastering}, supporting platforms such as \href{https://ethereum.org/}{Ethereum}, \href{https://polygon.technology/}{Polygon}, \href{https://arbitrum.io/}{Arbitrum}, \href{https://fantom.foundation/}{Fantom}, and \href{https://www.optimism.io/}{Optimism}, which are object of our study. Its distinct features \cite{dannen2017solidity}, including gas optimization \cite{li2021gas}, strict state management, and security constraints, introduce significant challenges for code analysis and quality assessment \cite{wohrer2018smart}. Traditional software metrics \cite{fenton1999software}, though effective for general-purpose programming languages, are often inadequate for evaluating Solidity’s unique properties, such as gas consumption, state variable organization, and access control mechanisms~\cite{destefanis2018smart, dhillon2021unpacking}.

Micro-patterns, defined as small and recognizable programming constructs that reflect intentional design decisions, were first introduced by Gil and Maman for Java~\cite{gil2005micro}. In their study, Gil and Maman demonstrated that 75\% of the classes in a Java system match at least one micro-pattern from their defined catalog. Unlike conventional software metrics, micro-patterns offer structured insights into code quality and evolution by capturing common structural and behavioral elements.
Building on this concept, we propose a novel framework for identifying and analyzing micro-patterns in Solidity code. 
We introduce a catalog of Solidity-specific micro-patterns that encapsulate its distinct programming characteristics. 
This work is the first systematic effort to apply micro-patterns to Solidity, providing a structured approach to evaluating smart contract.
In detail, we address the following research questions:

{\textbf{(RQ1) What micro-patterns can be identified in Solidity code, and how can they be systematically detected?}} We define a catalog of recurring micro-patterns in Solidity language and develop a framework for their automated detection.

{\textbf{(RQ2) To what extent are micro-patterns adopted in smart contract development?}} We analyze the prevalence of micro-patterns in a large dataset of smart contracts to understand their role in development practices.

{\textbf{(RQ3) How do micro-pattern frequencies vary across different blockchain platforms?}} We examine how micro-pattern adoption differs across platforms, highlighting variations and commonalities in development strategies.

{\textbf{(RQ4) Which relationships exist between different micro-patterns, and how are they spread?}} Through correlation and exclusivity analysis, we investigate how micro-patterns co-occur and interact in Solidity development.

Our contributions are as follows. First, we introduce a catalog of 18 micro-patterns grouped into five categories: \textit{security, functional, optimization, interaction, and feedback}. 
Each micro-pattern is formally defined to enable automated detection and to reflect critical aspects of Solidity development.
Second, we develop a micro-pattern detection framework with 93\% success rate across different Solidity versions. 
Our framework enables large-scale analysis of contracts on any Solidity-compatible blockchain and generates comprehensive metrics including pattern frequency, coverage rates, and cross-pattern correlation measures.
Third, we conduct an empirical study of over 23258 smart contracts from the five most popular blockchains. Our frequency analysis shows that 99\% of contracts exhibit at least one micro-pattern, with an average of 2.76 micro-patterns per contract.
The correlation analysis reveals that micro-patterns represent independent design choices, with predominantly weak relationships ($\phi$ < 0.15) between pattern pairs. We also perform a cross-platform comparison identifying significant variations in pattern adoption, particularly in optimization and security, reflecting platform-specific development practices and constraints.
All materials presented in this paper, including the micro-pattern detection framework, datasets, and analysis scripts, are fully replicable and verifiable in our replication package \href{https://figshare.com/s/19597346b31f7f78901a}{\textbf{link}}.

The remainder of this paper is organized as follows. Section \ref{sec:related} reviews related work on micro-patterns and smart contract analysis. Section \ref{sec:analysis} introduces our catalog of 18 micro-patterns and identification method, while Section \ref{sec:framework} describes our framework for automated detection. Section \ref{sec:datacollection} covers data collection and preparation, followed by Section \ref{sec:results} with the empirical evaluation. Section \ref{sec:threats} discusses threats to validity and Section \ref{sec:conclusion} concludes the paper.

\section{Related Work}
\label{sec:related}

Research on micro-patterns has evolved from their initial definition for Java to applications in code quality analysis. 
We survey these developments alongside existing approaches to smart contract analysis to contextualize the current understanding of code structures in both traditional and blockchain environments.

\paragraph{\textbf{Evolution of Micro-Pattern Analysis}}

Micro-patterns are advanced through several key developments in detection and classification methods. Arcelli et al. \cite{arcelli2010metrics} redefined micro-patterns in terms of number of methods (NOM) and number of attributes (NOA) of a class. 
The relationship between micro-patterns and programming practices \cite{concas2013micro}, and code quality emerged as another important research direction. Kim et al. \cite{kim2006micro} examined micro-pattern evolution across software versions, identifying specific evolution types associated with increased defect rates. Building on this work, Destefanis et al. \cite{destefanis2012micro} analyzed multiple Eclipse releases and found that classes not matching any micro-pattern showed higher fault rates than those exhibiting patterns. These findings about pattern presence shaped our method for evaluating Solidity code quality.

\paragraph{\textbf{Micro-Patterns for Quality and Security Assessment}}

The application of micro-patterns has extended into security analysis, with researchers demonstrating their value for vulnerability detection. Sultana et al. \cite{sultana2017evaluating,sultana2019study} identified correlations between certain patterns and security vulnerabilities in Apache Tomcat, showing how pattern analysis could enhance traditional vulnerability prediction methods that relied solely on software metrics. This security-focused application aligns with our work, as smart contracts face security challenges that can benefit from pattern-based analysis.
Codabux et al. \cite{codabux2017relationship} further demonstrated the value of combining pattern analysis with other techniques by investigating relationships between traceable patterns and code smells. Their work indicated that certain patterns may signal design issues, a finding that influenced our identification of problematic patterns in smart contracts. While these studies show the utility of micro-patterns in code analysis, their findings stem from object-oriented programming concepts specific to traditional software environments.

\paragraph{\textbf{Smart Contract Analysis Techniques}}

Current smart contract analysis research has focused primarily on static analysis of code structure and dynamic analysis of runtime properties. Ghaleb et al. \cite{ghaleb2020effective} evaluated six static analysis tools through systematic bug injection, revealing limitations in detecting known vulnerability patterns. Their work demonstrates gaps in current tools, but focuses on vulnerability detection rather than identifying intentional design decisions in smart contract development practices.
Static analysis tools employ various approaches. SmartCheck uses pattern matching through XPath expressions \cite{tikhomirov2018smartcheck}, while Slither \cite{feist2019slither} exploits AST analysis to discover a predefined set of vulnerabilities.
Instead, other solutions such as Mythril \cite{durieux2020empirical} and Manticore apply symbolic execution \cite{mossberg2019manticore} to reproduce smart contract execution and found possible issues. As highlighted by Vidal et al. \cite{vidal2024vulnerability}, while these tools can detect predefined vulnerabilities, they struggle with more complex scenarios that involve multiple contracts.

Research has extensively explored formal verification for smart contracts, with formal methods being particularly effective despite limited adoption \cite{singh2020blockchain,8763832, 10.1145/3464421}. The focus has been primarily on functionality verification, with fewer works addressing security specifically. While these works provide correctness guarantees, they do not address the identification of common programming patterns that could guide better development practices. As noted by Atzei et al. \cite{atzei2017survey}, there remains a gap between theoretical security properties and practical development patterns.
Traditional software metrics have been adapted for smart contract evaluation, but often fail to capture blockchain concerns like gas optimization and state management \cite{vacca2021systematic}. Recent studies identify authorization and authentication as major security risks in Ethereum smart contracts, especially those with external dependencies \cite{chen2020survey}. Destefanis et al.'s work \cite{destefanis2012micro} on detecting problematic patterns through metrics suggests approaches that could be modified for smart contracts.

\paragraph{\textbf{Research Gap and Our Contributions}}

The existing literature demonstrates both the value of micro-pattern analysis in traditional software and the need for specialized evaluation techniques for smart contracts. Current approaches apply patterns that do not translate well to blockchain environments, or analyze smart contracts without systematic pattern recognition. Some studies examine contract design aspects in isolation, missing the broader context of how patterns interact. 
We build upon Gil and Maman's \cite{gil2005micro} approach to micro-pattern definition while addressing the distinct characteristics of Solidity programming. Unlike previous adaptations of object-oriented metrics, our approach targets Solidity-specific concerns such as gas costs optimization and security.

\begin{table*}
\caption{Solidity Micro-Patterns Catalog}
    \centering
    \resizebox{\textwidth}{!}{
    \begin{tabular}{l|l|l|l}
        \toprule
        \textbf{Category} & \textbf{Name} & \textbf{Description} & \textbf{Entity Types} \\
        \midrule
        \multirow{4}{*}{\textbf{Security}} 
        & \textbf{Ownable} & An entity that maintains an owner's address and restricts access to specific functions using modifiers. & Contract \\ \cline{2-3}
        & \textbf{Stoppable} & An entity with a toggable state, so it can be paused, resumed, or permanently stopped if necessary. & Contract \\ \cline{2-3}
        & \textbf{Pull Payment} & A payment model where recipients withdraw funds themselves instead of receiving from the contract. & Contract \\ \cline{2-3}
        & \textbf{Reentrancy Guard} & A modifier that prevents reentrancy attacks by ensuring state updates occur before any external calls. & Contract \\
        \cline{2-3} \hline
        \multirow{4}{*}{\textbf{Functional}} & \textbf{Payable} & An entity that includes \texttt{fallback()} and \texttt{receive()} functions to enable the receipt of funds. & Contract \\ 
        \cline{2-3}
        & \textbf{Borrower} & An entity that utilizes external libraries to perform predefined operations on specific data types. & Contract \\ 
        \cline{2-3}
        & \textbf{Implementer} & An entity that implements all functions inherited from interfaces or abstract contracts. & Contract \\ \cline{2-3}
        & \textbf{Modifier Usage} & An entity that leverages reusable modifiers to enforce common conditions across multiple functions. & Contract, Library \\ \hline
        \multirow{3}{*}{\textbf{Optimization}} & \textbf{Storage Saver} & An entity that minimizes storage costs by efficiently arrange state variables into fewest storage slots. & Contract \\ \cline{2-3}
        & \textbf{Reader} & An entity where all functions are defined as \texttt{view}, ensuring they do not modify the contract state. & Contract, Interface, Library \\ \cline{2-3}
        & \textbf{Operator} & An entity where all functions are defined as \texttt{pure}, so they do not read or modify any state variable. & Contract, Interface, Library \\ \hline
        \multirow{3}{*}{\textbf{Interaction}} & \textbf{Provider} & An entity where all functions are \texttt{external} so that only external deployed contracts can invoke them. & Contract, Interface, Library \\ \cline{2-3}
        & \textbf{Supporter} & An entity where all functions are \texttt{internal} so that only the contract and its inheritors can use them. & Contract, Interface, Library \\ \cline{2-3}
        & \textbf{Delegator} & An entity that delegates operations by invoking functions in another deployed contract. & Contract, Library \\ \hline
        \multirow{4}{*}{\textbf{Feedback}}
        & \textbf{Named Return} & An entity where return values are explicitly assigned names in the function definition. & Contract, Interface, Library \\ \cline{2-3}
        & \textbf{Returnless} & An entity that does not return any values from its functions. & Contract, Interface, Library \\ \cline{2-3}
        & \textbf{Emitter} & An entity where every function execution emits at least one event. & Contract, Library \\ \cline{2-3}
        & \textbf{Muted} & An entity where no events are emitted during the execution of its functions. & Contract, Library \\
        \bottomrule
    \end{tabular}
    }
    \label{tab:pattern-catalog}
\end{table*}

\section{Micro-Pattern Analysis}\label{sec:analysis}

This section presents our method for identifying and categorizing micro-patterns in Solidity, addressing RQ1's focus on recognizable development micro-patterns.
We begin by describing the identification process, followed by an organized catalog based on functional roles in Solidity development. Finally, we introduce our framework for automated detection.

\subsection{Micro-Pattern Identification Process}

Our method for identifying micro-patterns in Solidity builds upon Gil and Maman's~\cite{gil2005micro} systematic process while adapting it for blockchain-specific development practices\footnote{\url{https://docs.soliditylang.org/en/latest/style-guide.html}}. %
First, we analyzed how Gil and Maman's Java micro-patterns could be translated to Solidity's domain. We examined each micro-pattern in their catalog and evaluated whether it could meaningfully capture Solidity development practices. This revealed that while some micro-patterns (like \textit{Implementor}) could transfer directly, blockchain's characteristics required new patterns. 

Second, we conducted an iterative pattern discovery process:
\begin{enumerate}
    \item We examined Solidity's distinctive features (e.g., modifiers, view/pure functions, event emissions) and considered meaningful restrictions on their usage;
    \item For each potential micro-pattern, we analyzed its impact in addressing specific concerns like ``what are the implications of state variable arrangements on gas costs?'';
    \item We implemented initial micro-pattern definitions for applying them to public contract repositories and evaluate their behaviours;
    \item Through manual code inspection of matched contracts, we refined definitions, merged similar micro-patterns, and discarded those that did not capture design decisions;
    \item When automatic pattern detection revealed clusters of similar but not identical implementations, we analyzed whether these represented meaningful variants requiring pattern refinement.
\end{enumerate}

Finally, we validated all candidate micro-patterns against three key criteria defined in the following:

\paragraph{\textbf{Design Intent.}}
A micro-pattern must capture a meaningful architectural or functional aspect in Solidity development design practices. For instance, the \textit{Storage Saver} micro-pattern (Table \ref{tab:pattern-catalog}) reflects an optimization strategy where developers arrange state variables to minimize storage slots, reducing gas costs. This distinguishes meaningful micro-patterns from arbitrary structures. For example, a contract that increments a counter or uses simple conditionals may create recognizable patterns, but these do not represent deliberate design decisions.
\paragraph{\textbf{Mechanical Recognizability.}}
A micro-pattern must be expressible as a condition that can be verified through static analysis \cite{gil2005micro}. For example, verifying whether all state-modifying functions implement a reentrancy guard can be achieved by analyzing the presence of specific modifiers and function calls in the code. We formalized each micro-pattern using first-order logic (FOL) specifications, defining precise conditions on contract components, methods, and modifiers (see Appendix A: Table 4). This formal definition ensures automatic and unambiguous detection.
\paragraph{\textbf{Empirical Validation.}}
A micro-pattern must be observable in deployed smart contracts across multiple blockchain platforms. We validated each micro-pattern's adoption through analysis of contracts, ensuring that it represents a recurring development practice rather than an isolated case. This empirical validation distinguishes meaningful micro-patterns implemented with a specific scope from coincidental code structures.\\

\begin{figure*}[ht]
    \centering
    \includegraphics[width=1\linewidth]{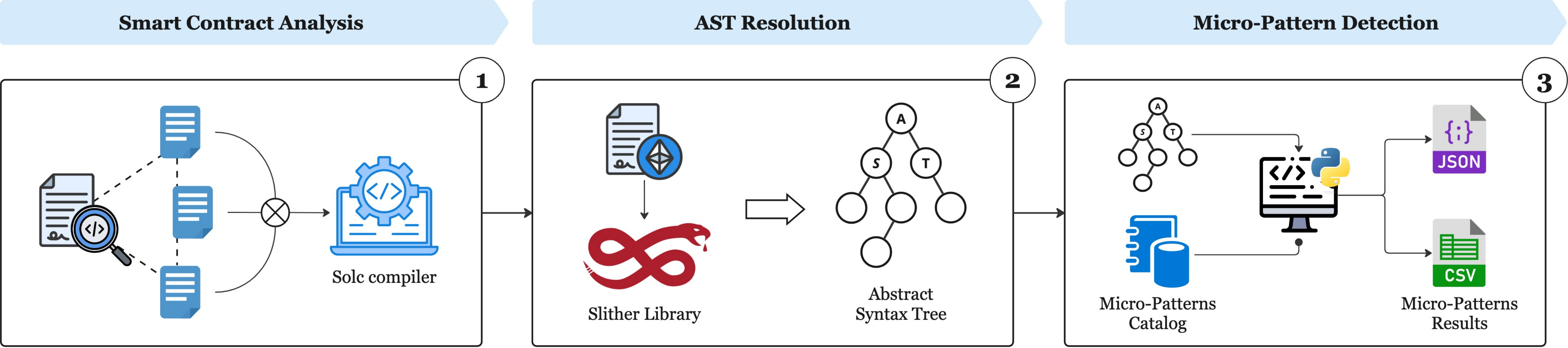}
    \caption{Micro-Pattern Detection Framework}
    \label{fig:tool}
\end{figure*}

Following these guidelines, our method balances automated detection with the need to capture meaningful design choices. The combination of mechanical recognizability, design intent, and empirical evidence ensures that the identified micro-patterns represent genuine development practices addressing challenges in Solidity smart contracts. Patterns that were infrequent, highly contract-specific, or not mechanically detectable were excluded.

\subsection{Micro-Patterns Catalog}

After following the identification process, we defined 18 micro-patterns reported in Table \ref{tab:pattern-catalog}. 
While additional ones may exist, this catalog contains the most frequent and practically relevant ones based on our criteria.
The categorization reflects primary concerns in smart contract development, including \textit{security}, \textit{gas costs optimization}, and \textit{cross-contract interactions}.

\textbf{Security Patterns.} The immutable nature of deployed smart contracts requires built-in security measures. Following Gil and Maman's observation that micro-patterns can capture language-specific features, these structures reflect different approaches to address unique security requirements of smart contracts compared to traditional software. For example, the \textit{Ownable} micro-pattern implements single-owner authorization, restricting functions execution. Instead, \textit{Pull Payment} and \textit{Reentrancy Guard} implement protection mechanisms that, once deployed, cannot be modified.

\textbf{Functional Patterns.} Similar to how Java micro-patterns capture core operational features, this category encompasses key smart contract behaviors. These include micro-patterns such as \textit{Pull Payment} that enables smart contracts to receive funds and modularity mechanisms such as \textit{Borrower} and \textit{Implementer}. They demonstrate how to leverage Solidity features to create maintainable and extensible contracts.

\textbf{Optimization Patterns.} The gas-cost model of the Ethereum Virtual Machine creates unique optimization requirements. These micro-patterns focus on improving resource efficiency through techniques like the \textit{Storage Saver} for minimizing storage costs. This category has no direct parallel in traditional micro-patterns, reflecting blockchain-specific concerns.

\textbf{Interaction Patterns.} Smart contracts rarely operate in isolation. Interaction micro-patterns govern contract communication and functionality delegation. Unlike object-oriented micro-patterns that focus on class relationships, these structures define standard approaches for inter-contract communication in the decentralized blockchain ecosystem.

\textbf{Feedback Patterns.} The final category addresses how smart contracts communicate state changes and execution results. Through event emission and return values, these micro-patterns define how smart contracts interact with external systems. This category reflects the need for observable behavior in decentralized applications.

The organization into these categories emerges from both the intrinsic properties of the Solidity language and the practical requirements of smart contract development.

\begin{tcolorbox}[right=0.1cm,left=0.1cm,top=0.1cm,bottom=0.1cm]
    \textbf{\textit{(RQ1) What micro-patterns can be identified in Solidity code, and how can they be systematically detected?}}
    Through analysis of Solidity features and development practices, we identified 18 micro-patterns across five categories. Each micro-pattern represents a recognizable and reproducible development practice. The catalog provides the foundation for understanding and implementing common smart contract functionalities, with micro-patterns applicable across different blockchain platforms and contract types.
\end{tcolorbox}

\begin{figure*}[ht!]
 \centering
 \includegraphics[width=\textwidth]{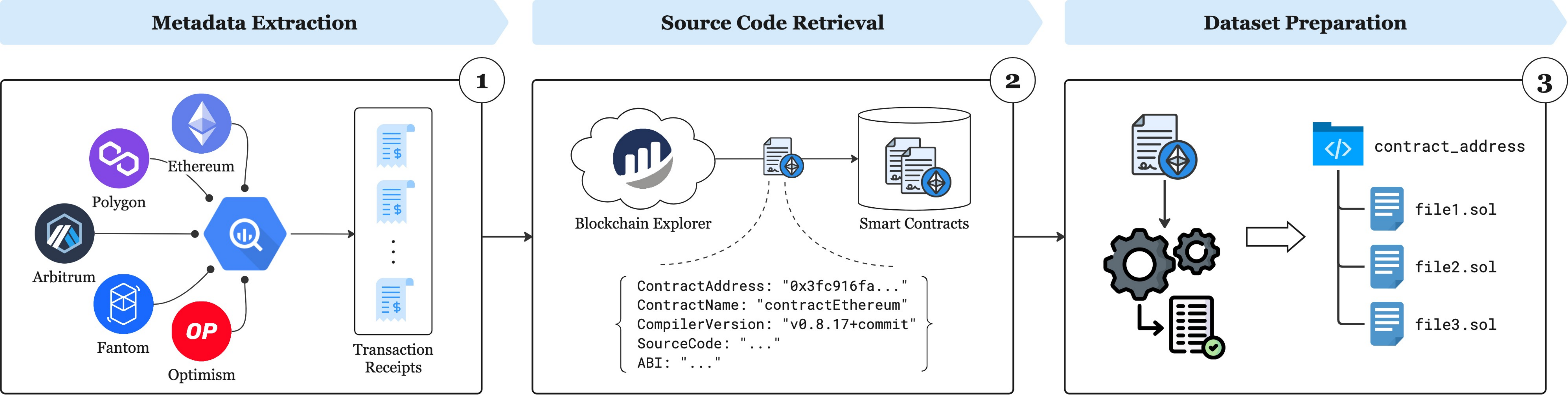}
 \caption{Data Collection Method}
 \Description{}
 \label{fig:methodology}
\end{figure*}

\section{Micro-Pattern Detection Framework}
\label{sec:framework}
We implemented a Python-based framework, whose workflow is depicted in Figure \ref{fig:tool}, to analyze Solidity files and detect micro-patterns in the code. It supports both single-file contracts and those organized across multiple sources, manages compiler version resolution for different Solidity compiler versions, and handles dependency management by resolving import paths and ensuring the availability of required dependencies.

The detection process begins by scanning the input paths to identify all Solidity files and their associated dependencies, as shown in Figure \ref{fig:tool} \textbf{(step 1)}. Once the files are identified, the tool analyzes them to extract the latest compatible Solidity compiler versions required for each contract, resolving the versions based on pragma directives. To optimize the compilation process, the tool orders the files by compiler version, reducing unnecessary compiler switches and improving efficiency.

After determining the compilation order, the process continues as presented in Figure \ref{fig:tool} \textbf{(step 2)}, where the framework uses the Slither library\footnote{Slither library: \url{https://github.com/crytic/slither/}} \cite{feist2019slither} to compile each smart contract and extract the Abstract Syntax Tree (AST). 
The framework implements our formal First-order-logic (FOL) specifications through AST traversal operations. Each pattern detection algorithm systematically checks AST nodes according to the conditions defined in our FOL specifications. The implementation's correctness is validated through testing against known pattern examples.

With the AST representation of the code, the framework traverses the structure to identify occurrences of predefined micro-patterns in each entity, which may include contracts, abstract contracts, interfaces, and libraries. The detection process not only analyzes individual entities but also considers inheritance hierarchies, modifiers, and external dependencies to ensure accurate results. Dependencies such as imported contracts and external libraries are automatically resolved and analyzed alongside the provided contracts, ensuring that all necessary components are available for micro-pattern detection.

For each analyzed entity, the tool collects and outputs detailed information as in Figure \ref{fig:tool} \textbf{(step 3)}, including the contract's name, file path, compiler version, type (i.e., contract, abstract contract, interface, or library), and whether each micro-pattern is present. The results are then exported in the user-specified format, enabling further analysis of micro-pattern prevalence and distribution across the analyzed smart contracts.

Our framework achieved a 93\% success rate in processing 23258 out of 25000 verified smart contracts. 
The unsuccessful cases were primarily due to dependency resolution issues and import path remapping problems in the downloaded smart contract sources, rather than limitations in the pattern detection logic itself.
Overall, this framework ensures accurate and efficient detection of micro-patterns while adapting to the diverse structures and requirements of Solidity projects.

\section{Data Collection}
\label{sec:datacollection}

To empirically validate the micro-pattern catalog and address our research questions, we follow the method in Figure \ref{fig:methodology} to collect a large dataset of verified smart contracts, including both metadata and source code.
Verified smart contracts are those whose source code is available and validated in blockchain explorers\footnote{Verified smart contracts list: \url{https://etherscan.io/contractsVerified}} - web platforms that index and display blockchain transaction data. Our detection framework, described in Section \ref{sec:framework}, requires this source code to accurately identify micro-pattern occurrences through static analysis. This dataset supports our empirical evaluation of micro-pattern adoption (RQ2), distribution across different blockchain platforms (RQ3), and relationships between micro-patterns (RQ4).

\subsection{Metadata Extraction}

We extracted all transaction receipts, from the genesis block up to 31st December 2024,
from five blockchain networks: Ethereum, Polygon, Arbitrum, Fantom, and Optimism.
These networks were selected based on their representation of deployed contracts and availability in Google BigQuery Web3 public datasets\footnote{Google Web3 Datasets: \url{https://cloud.google.com/application/web3/discover/}}.

As in Figure \ref{fig:methodology} (step 1), we retrieved contract metadata from transaction receipts with non-null \texttt{contract\_address} fields on BigQuery database, indicating successful deployment of a smart contract.
For each smart contract, we collect the following information: contract address, creator address, block timestamp, block number, and transaction hash.
Contract and creator addresses allow tracking contract origins and ownership micro-patterns. Block number and timestamp, instead, provide temporal context for micro-pattern evolution analysis. Finally, transaction hash serves as a unique identifier and enable verification of deployment details.

\begin{table}[h!]
\caption{Blockchain Entities Statistics}
\centering
 \resizebox{1\columnwidth}{!}{
    \begin{tabular}{c|c|cccc|r}
        \toprule
        \multirow{2}{*}{\textbf{Blockchain}} & \textbf{Smart} & \multicolumn{5}{c}{\textbf{Entities}} \\
        \cline{3-7}
        & \textbf{Contracts} & \textbf{Contract} & \textbf{Abstract C.} & \textbf{Interface} & \textbf{Library} & 
        \textbf{Total} \\
        \midrule
        Ethereum & 4.855 & 10.097 & 10.159 & 20.092 & 10.712 & 51.060 \\
        Polygon & 4.900 & 11.666 & 16.664 & 23.341 & 12.092 & 63.763 \\
        Arbitrum & 4.415 & 34.935 & 24.113 & 46.382 & 25.328 & 130.758 \\
        Fantom & 4.871 & 11.154 & 12.437 & 30.657 & 14.256 & 68.504 \\
        Optimism & 4.217 & 7.708 & 10.039 & 18.732 & 11.243 & 47.722 \\
        \midrule
        \textbf{Total} & 23.258 & 75.560 & 73.412 & 139.204 & 73.631 & 361.807 \\
        \bottomrule
    \end{tabular}}  
    \label{tab:blockchain_statistics}
\end{table}

\begin{figure*}[t!]
\begin{tcolorbox}[
    colback=lightBg,
    colframe=borderColor,
    boxrule=0.5pt,
    arc=0mm]
\begin{minipage}{0.59\textwidth}
\begin{align*}
\textcolor{mainBlue}{E} &= \{\,e \mid\; \text{entity } e \in \mathrm{Dataset} \\[1pt]
&\;\wedge\; \textcolor{emphColor}{\text{type}}(e) \in \{\texttt{contract},\,\texttt{library},\,\mathtt{interface}\}\} 
\\[3pt]
\textcolor{mainBlue}{MP} &\;=\; \{\, mp \;\mid\; mp \text{ is one of the defined micro-patterns} \}
\\[3pt]
\textcolor{mainBlue}{\text{ValidTypes}}(mp) 
&\;\subseteq\;
\{\mathtt{contract},\,\mathtt{interface},\,\mathtt{library}\}
\\[5pt]
\textcolor{mainBlue}{M}(mp, e) &=
\begin{cases}
    1, & \text{if } \text{type}(e)\in \text{ValidTypes}(mp)
         \text{ and } e \text{ satisfies } mp,\\
    0, & \text{otherwise}.
\end{cases}
\end{align*}
\end{minipage}
\hfill
\begin{minipage}{0.39\textwidth}
\begin{align*}
\textcolor{mainBlue}{\text{Frequency}}(mp) &= \sum_{\,e \,\in\, E(mp)} M(mp, e)
\\[12pt]
\textcolor{mainBlue}{\text{Coverage}}(mp) &= \frac{\text{Frequency}(mp)}{\lvert E(mp)\rvert}
\\[12pt]
\textcolor{mainBlue}{\text{Prevalence}}(mp) &= \frac{\text{Frequency}(mp)}{\sum_{\,mp' \in MP\,} \text{Frequency}(mp')}
\end{align*}
\end{minipage}
\end{tcolorbox}
\caption{Analysis framework. Left: \textit{E} captures entities (contracts, interfaces, libraries) from these projects, \textit{MP} defines micro-patterns, \textit{ValidTypes} maps patterns to applicable entity types, and \textit{M} is the matching function. Right: Core metrics measure pattern occurrence (\textit{Frequency}), adoption rate among eligible entities (\textit{Coverage}), and relative dominance (\textit{Prevalence}).}
\label{fig:definitions}
\end{figure*}

\subsection{Source Code Retrieval}

From the extracted metadata, we collected an adequate number of verified smart contracts per platform, obtaining their source code through blockchain explorers (Etherscan, Polygonscan, Arbiscan, Ftmscan, and Optimistic)\footnote{Explorer URLs: \url{https://etherscan.io/}, \url{https://polygonscan.com/}, \url{https://arbiscan.io/}, \url{https://ftmscan.com/}, \url{https://optimistic.etherscan.io/}} as shown in Figure \ref{fig:methodology} (step 2). 
The sample size must provide statistical power for analyzing micro-pattern prevalence (RQ2), cross-blockchain comparisons (RQ3), and correlation analysis (RQ4).

To determine the minimum required sample size of our dataset, we conducted a statistical power analysis for chi-square tests used in comparing micro-pattern distributions across platforms. With five blockchains, our analysis involves ten pairwise comparisons ($\frac{5(5-1)}{2}=10$), necessitating a Bonferroni-corrected significance level of $\alpha = 0.005$ ($0.05/10$) \cite{weisstein2004bonferroni}.
Power analysis indicated that detecting small effect sizes ($w = 0.1$) requires a minimum of 1332 smart contracts per blockchain. We chose to collect at least 4000 smart contracts for each blockchain to enhance the statistical validity of our analysis in several ways. First, this larger sample size enables detection of effects smaller than $w = 0.1$, allowing identification of subtle variations in micro-pattern adoption across platforms. Second, when examining relationships among 18 micro-patterns (RQ4), the increased sample size reduces variance in correlation estimates, particularly for weak associations. Finally, given the heterogeneous nature of smart contract deployment across blockchain platforms, having at least 4000 smart contracts per platform provides more reliable coverage of diverse implementation patterns.

As detailed in Table \ref{tab:blockchain_statistics}, our dataset encompasses both standalone smart contracts and decentralized applications (DApps). Standalone smart contracts implement specific functionalities such as token management or voting systems, while DApps comprise multiple interconnected smart contracts with complex state management requirements. This architectural diversity strengthens our analysis of micro-pattern relationships (RQ4) and design decisions (RQ2).
Moreover, the dataset includes smart contracts implemented with different Solidity versions across the five blockchain platforms, ranging from compiler version 0.4.x to 0.8.x.
We organized these in blockchain-specific SQL databases to facilitate efficient micro-pattern detection and cross-chain analysis for RQ3, enabling both granular and comparative studies of micro-pattern adoption.

\subsection{Dataset Preparation}

After retrieving the smart contracts' source code, we processed and organized the files to preserve their dependency relationships, as illustrated in Figure \ref{fig:methodology} (step 3).
This preparation phase addressed three key challenges in Solidity compilation: version compatibility, dependency resolution, and multi-file project organization.
We processed each smart contract differently based on its source code structure.
For single-file smart contracts, we maintain their original structure while verifying version compatibility.
On the other hand, we implement a systematic dependency resolution mechanism for multi-file smart contracts that reconstructs the project hierarchy.
This process analyzes Solidity import statements and remaps paths to maintain the smart contract's intended structure, handling both direct dependencies and custom library paths.

Based on this, we created a structured dataset for each of the selected blockchain.
We stored the processed smart contracts in separate directories, each containing a complete and self-contained project. 
The directory structure that preserves the relationships between smart contract components while resolving potential conflicts in import paths and library references.
This organization ensures that when the Abstract Syntax Tree (AST) is generated for micro-pattern detection (Section \ref{sec:framework}), all inheritance hierarchies and external dependencies are properly resolved for the analysis. 
The complete dataset, including all processed smart contracts and their source code, is available in our replication package \href{https://figshare.com/s/19597346b31f7f78901a}{\textbf{link}}.

\section{Empirical Evaluation}\label{sec:results}

This section presents the empirical evaluation of micro-patterns in Solidity language, analyzing pattern adoption rates (RQ2), cross-platform variations (RQ3), and pattern relationships (RQ4) across smart contracts in our dataset. 
This evaluation demonstrates the prevalence of identified micro-patterns in real-world smart contract development, quantifies relationships between different micro-patterns, and reveals how developers adopt them across different blockchain platforms. Through systematic analysis of our dataset, we examine both the individual characteristics of each micro-pattern and their collective impact on smart contract development.

\subsection{Pattern Adoption Analysis (RQ2)}

\begin{table*}[t!]
\caption{Metrics for Micro-Patterns across Blockchains with Coverage Statistics. For each blockchain, \colorbox{blue!30}{blue-shaded cells} indicate micro-patterns with the highest adoption rates, while \colorbox{cyan!25}{azure-shaded cells} micro-patterns with the lowest adoption.}
\centering
\resizebox{\textwidth}{!}{
\begin{tabular}{l|l|rrr|rrr|rrr|rrr|rrr|r|r}
    \toprule
    \multirow{2}{*}{\textbf{Category}} & \multirow{2}{*}{\textbf{Micro-Pattern}}
    & \multicolumn{3}{c|}{\textbf{Ethereum}} 
    & \multicolumn{3}{c|}{\textbf{Polygon}} 
    & \multicolumn{3}{c|}{\textbf{Arbitrum}} 
    & \multicolumn{3}{c|}{\textbf{Fantom}} 
    & \multicolumn{3}{c|}{\textbf{Optimism}}
    & \multicolumn{2}{c}{\textbf{Total Cov. Stats (\%)}} \\
    \cline{3-19}
     &  & \textbf{Freq.} & \textbf{Cov. (\%)} & \textbf{Prev. (\%)} 
         & \textbf{Freq.} & \textbf{Cov. (\%)} & \textbf{Prev. (\%)} 
        & \textbf{Freq.} & \textbf{Cov. (\%)} & \textbf{Prev. (\%)} 
        & \textbf{Freq.} & \textbf{Cov. (\%)} & \textbf{Prev. (\%)} 
        & \textbf{Freq.} & \textbf{Cov. (\%)} & \textbf{Prev. (\%)}
        & \textbf{Mean$\pm\sigma$} & \textbf{Median} \\
    \midrule
    \multirow{4}{*}{\textbf{Security}} 
        & Ownable & 269 & 1.33 & 0.20
        & 743 & 2.62 & 0.41 
        & 446 & 0.76 & 0.11 
        & 947 & 4.01 & 0.53 
        & 98 & 0.55 & 0.07 
        & 1.85$\pm$1.30 & 1.33 \\
        & Stoppable & 448 & 2.21 & 0.33 
        & 1328 & 2.25 & 0.34
        & 1328 & 2.25 & 0.26 
        & 368 & 1.56 & 0.21
        & 436 & 2.46 & 0.33 
        & 2.15$\pm$0.31 & 2.25 \\
        & Pull Payment & 61 & 0.30 & 0.04 
        & 32 & 0.11 & 0.02
        & \cellcolor{cyan!25} 47 & \cellcolor{cyan!25} 0.08 & \cellcolor{cyan!25} 0.01
        & \cellcolor{cyan!25} 21 & \cellcolor{cyan!25} 0.09 & \cellcolor{cyan!25} 0.01
        & 32 & 0.18 & 0.02 
        & 0.15$\pm$0.08 & 0.11 \\
        & Reentrancy Guard & \cellcolor{cyan!25} 39 & \cellcolor{cyan!25} 0.19 & \cellcolor{cyan!25} 0.03 
        & \cellcolor{cyan!25} 29 & \cellcolor{cyan!25} 0.10 & \cellcolor{cyan!25} 0.02
        & 119 & 0.20 & 0.03
        & 101 & 0.43 & 0.06
        & \cellcolor{cyan!25} 8 & \cellcolor{cyan!25} 0.05 & \cellcolor{cyan!25} 0.01 
        & 0.19$\pm$0.13 & 0.19 \\
    \midrule
    \multirow{4}{*}{\textbf{Functional}} 
        & Payable & 1040 & 5.13 & 0.76
        & 2452 & 8.66 & 1.37 
        & 2153 & 3.65 & 0.55 
        & 3112 & 13.19 & 1.75 
        & 2448 & 13.79 & 1.86 
        & 8.88$\pm$4.10 & 8.66 \\
        & Borrower & 7961 & 39.30 & 5.80 
        & 11818 & 41.72 & 6.60
        & 37018 & 62.69 & 9.43 
        & 8225 & 34.86 & 4.62 
        & 8418 & 47.43 & 6.40 
        & 45.20$\pm$9.64 & 41.72 \\
        & Implementer & 7942 & 39.21 & 5.79 
        & 12233 & 43.18 & 6.83 
        & 35219 & 59.64 & 8.97 
        & 10363 & 43.93 & 5.83
        & 8111 & 45.70 & 6.17 
        & 46.33$\pm$6.98 & 43.93 \\
        & Modifier Usage & 10417 & 33.64 & 7.60 
        & 14211 & 35.16 & 7.94 
        & 31114 & 36.88 & 7.93 
        & 10194 & 26.93 & 5.73
        & 7879 & 27.18 & 5.99 
        & 31.96$\pm$4.13 & 33.64 \\
    \midrule
    \multirow{3}{*}{\textbf{Optimization}} 
        & Storage Saver & 16301 & 80.47 & 11.89 
        & \cellcolor{blue!30} 25969 & \cellcolor{blue!30} 91.67 & \cellcolor{blue!30} 14.50 
        & 42195 & 71.46 & 10.75
        & 21854 & 92.64 & 12.29 
        & 15413 & 86.85 & 11.72 
        & 84.62$\pm$7.87 & 86.85 \\
        & Reader & 6857 & 13.43 & 5.00 
        & 9956 & 15.61 & 5.56 
        & 13473 & 10.30 & 3.43 
        & 8782 & 12.82 & 4.94
        & 6391 & 13.39 & 4.86 
        & 13.11$\pm$1.70 & 13.39 \\
        & Operator & 4743 & 9.29 & 3.46 
        & 4551 & 7.14 & 2.54
        & 9184 & 7.02 & 2.34 
        & 4066 & 5.94 & 2.29
        & 3945 & 8.27 & 3.00 
        & 7.53$\pm$1.15 & 7.14 \\
    \midrule
    \multirow{3}{*}{\textbf{Interaction}} 
        & Provider & 18422 & 36.08 & 13.43
        & 20386 & 31.97 & 11.38 
        & 44519 & 34.05 & 11.34 
        & \cellcolor{blue!30} 26128 & \cellcolor{blue!30} 38.14 & \cellcolor{blue!30} 14.69 
        & 16830 & 35.27 & 12.80 
        & 35.10$\pm$2.06 & 35.27 \\
        & Supporter & 11123 & 21.78 & 8.11 
        & 11480 & 18.00 & 6.41 
        & 22751 & 17.40 & 5.80 
        & 12606 & 18.40 & 7.09
        & 8843 & 18.53 & 6.72 
        & 18.82$\pm$1.53 & 18.40 \\
        & Delegator & 13188 & 42.59 & 9.62 
        & 18789 & 46.48 & 10.49
        & \cellcolor{blue!30} 49541 & \cellcolor{blue!30} 58.71 & \cellcolor{blue!30} 12.62 
        & 17794 & 47.02 & 10.01
        & 14163 & 48.85 & 10.77 
        & 48.73$\pm$5.39 & 47.02 \\
    \midrule
    \multirow{4}{*}{\textbf{Feedback}}
        & Named Return & 16225 & 31.78 & 11.83 
        & 17754 & 27.84 & 9.91
        & 46541 & 35.59 & 11.86 
        & 23896 & 34.88 & 13.44 
        & 15780 & 33.07 & 12.00 
        & 32.63$\pm$2.75 & 33.07 \\
        & Returnless & 4850 & 9.50 & 3.54 
        & 7357 & 11.54 & 4.11 
        & 12611 & 9.64 & 3.21
        & 7574 & 11.06 & 4.26 
        & 5598 & 11.73 & 4.26 
        & 10.69$\pm$0.94 & 11.06 \\
        & Emitter & 191 & 0.62 & 0.14 
        & 121 & 0.30 & 0.07 
        & 287 & 0.34 & 0.07
        & 150 & 0.40 & 0.08 
        & 139 & 0.48 & 0.11 
        & 0.43$\pm$0.11 & 0.40 \\
        & Muted & \cellcolor{blue!30} 17075 & \cellcolor{blue!30} 55.14 & \cellcolor{blue!30} 12.45 
        & 20486 & 50.68 & 11.44 
        & 44036 & 52.19 & 11.22
        & 21661 & 57.23 & 12.18
        & \cellcolor{blue!30} 16965 & \cellcolor{blue!30} 58.52 & \cellcolor{blue!30} 12.90 
        & 54.75$\pm$2.95 & 55.14 \\
    \midrule
    \multicolumn{2}{c|}{\textbf{Total Coverage (\%)}} 
        & \multicolumn{3}{c|}{99.98} 
        & \multicolumn{3}{c|}{99.99} 
        & \multicolumn{3}{c|}{99.99} 
        & \multicolumn{3}{c|}{100.00} 
        & \multicolumn{3}{c|}{99.98}
        & \multicolumn{2}{c}{99.99$\pm$0.01} \\
    \bottomrule
\end{tabular}}
\label{tab:pattern-metrics}
\end{table*}

In the first part of the evaluation, we examined the extent of micro-pattern adoption in smart contract development. We analyzed statistical metrics to understand the frequency and distribution of micro-patterns in Solidity smart contracts. Figure~\ref{fig:definitions} formalizes our analysis framework: the set \textit{E} captures all code entities across our dataset, where each entity (of type \texttt{contract}, \texttt{interface}, or \texttt{library}) may implement one or more micro-patterns, with pattern eligibility defined by \textit{ValidTypes}; the matching function \textit{M} returns a binary value indicating whether an eligible entity implements a specific micro-pattern (1) or not (0).

We evaluated micro-pattern adoption using three complementary metrics. \textit{Frequency} counts the absolute number of entities matching a pattern. \textit{Coverage} normalizes this count by the number of eligible entities, revealing the percentage of potential implementations that adopt the pattern. \textit{Prevalence} contextualizes each pattern's adoption by showing its frequency relative to all pattern occurrences, indicating its relative importance in the ecosystem.

Table~\ref{tab:pattern-metrics} presents the adoption metrics across five major blockchain platforms. Our analysis revealed widespread adoption of micro-patterns in smart contract development, with projects implementing an average of 2.76 patterns (median: 2.00, $\sigma$: 1.43). The consistent adoption across platforms (means ranging from 2.60 to 3) suggests these patterns represent fundamental development practices.
\textit{The Storage Saver} pattern showed the highest coverage at 84.62\% ($\sigma$: 7.87\%). Its prevalence rates (10.75-14.50\%) align with what we would expect given the average number of patterns per contract, suggesting it is a fundamental pattern that is consistently implemented alongside other patterns rather than being used in isolation.

Approximately 28\% (5 out of 18) of our identified micro-patterns achieved wide adoption ($>$40\% coverage), distributed across functional (\textit{Borrower}: 45.20\%, \textit{Implementer}: 46.33\%), optimization (\textit{Storage Saver}: 84.62\%), interaction (\textit{Delegator}: 48.73\%), and feedback (\textit{Muted}: 54.75\%) categories.
Security-focused patterns showed lower but consistent adoption rates, with \textit{Reentrancy Guard} (mean: 0.19\%, $\sigma$: 0.13\%) and \textit{Pull Payment} (mean: 0.15\%, $\sigma$: 0.08\%) being selectively implemented. These stable rates across platforms indicated deliberate pattern selection for specific security requirements rather than oversight.
The varying adoption rates, from nearly universal optimization patterns to selective security patterns, demonstrated that our catalog successfully captured both foundational and specialized development practices in the Solidity ecosystem.

While our method considers pattern frequency as one indicator of meaningful development practices, we acknowledge that high adoption rates do not necessarily indicate best practices. For instance, the \textit{Muted} pattern's high coverage (54.75\%) might reflect insufficient logging practices rather than optimal design choices. Similarly, some infrequent patterns might represent important security practices that are underutilized. Therefore, our identification process focuses on capturing recurring structural choices in smart contract development, whether beneficial or potentially problematic, to provide a foundation for future research on pattern impact and quality outcomes.

\begin{tcolorbox}[right=0.1cm,left=0.1cm,top=0.1cm,bottom=0.1cm]
\textit{\textbf{(RQ2) To what extent are micro-patterns adopted in smart contract development?}}
Our identified micro-patterns are widely adopted, with projects implementing an average of 2.76 patterns (median: 2.00, $\sigma$: 1.43). We found that 28\% of them achieve high adoption rates ($>$40\% coverage), distributed across \textit{Functional, Optimization, Interaction}, and \textit{Feedback} categories, while others serve more specialized purposes. This distribution suggests these micro-patterns effectively capture both foundational and specialized smart contract development practices.
\end{tcolorbox}

\begin{figure*}[ht!]
\centering
\begin{minipage}{0.25\textwidth}
\centering
\caption*{(a) High correlation pairs ($\phi > 0.50$)}
\begin{tabular}{|l|c|}
\toprule
\textbf{Pattern Pair} & \textbf{$\phi$} \\
\midrule
Modifier Usage - Muted & 0.58 \\
Operator - Supporter & 0.52 \\
& \\
\bottomrule
\end{tabular}
\end{minipage}
\hfill
\begin{minipage}{0.26\textwidth}
\centering
\caption*{(b) Low correlation pairs ($\phi < 0.15$)}
\begin{tabular}{|l|c|}
\toprule
\textbf{Pattern Pair} & \textbf{$\phi$} \\
\midrule
Named Return - Ownable & 0.01 \\
Supporter - Emitter & 0.03 \\
Implementer - Emitter & 0.04 \\
\bottomrule
\end{tabular}
\end{minipage}
\hfill
\begin{minipage}{0.46\textwidth}
\centering
\caption*{(c) Mutual Exclusivity and Inclusivity by design}
\begin{tabular}{c|c}
\toprule
\textbf{Mutually Exclusive} & \textbf{Inclusive (Unidirectional)} \\
\midrule
Reader $\Leftrightarrow$ Operator & Ownable $\Rightarrow$ Modifier Usage \\
Emitter $\Leftrightarrow$ Muted & Stoppable $\Rightarrow$ Modifier Usage \\
Provider $\Leftrightarrow$ Supporter & \\
Multi Return $\Leftrightarrow$ Returnless & \\
Named Return $\Leftrightarrow$ Returnless & \\
\bottomrule
\end{tabular}
\end{minipage}
\caption{Pattern relationships analysis showing: (a) highly correlated pattern pairs, (b) low correlation pattern pairs (excluding zero correlations), and (c) patterns that are mutually exclusive or inclusive by design.}
\label{fig:pattern-relationships}
\end{figure*}

\subsection{Pattern Distribution Analysis (RQ3)}

To understand how blockchain environments influence design approaches, we analyzed micro-pattern distribution across platforms using both descriptive statistics and inferential analysis. With 4000+ smart contracts per blockchain platform (see Table~\ref{tab:blockchain_statistics}), we examined pattern adoption variations across platforms and then employed non-parametric statistical tests to validate observed differences.
We used non-parametric tests \cite{siegel1957nonparametric} because our data consisted of binary observations (presence/absence of micro-patterns) and not follow normal distribution assumptions required for parametric tests.

The descriptive analysis in Table ~\ref{tab:pattern-metrics} revealed distinct platform-specific adoption trends. Security micro-patterns showed notably different adoption rates, with Ethereum exhibiting higher coverage of security-focused patterns.
For instance, \textit{Reentrancy Guard} showed higher coverage on Ethereum (0.19\%) compared to Optimism (0.05\%), while \textit{Pull Payment} patterns had higher coverage on Ethereum (0.30\%) than other platforms (mean: 0.11\%). This aligned with Ethereum's role as the primary platform for high-value DeFi applications. Optimization micro-patterns demonstrated different trends: the \textit{Storage Saver} pattern achieved notably higher coverage on Polygon (91.67\%) and Fantom (92.64\%) compared to Arbitrum (71.46\%), suggesting different prioritization of optimization practices across platforms.

To identify meaningful differences in micro-pattern adoption between blockchains, we employed chi-square tests of independence with Bonferroni correction ($\alpha = 0.05/10 = 0.005$) for multiple comparisons. We complemented this with Cramer's V ($w$) threshold ($\geq$0.10) to focus on practically significant differences rather than just statistical significance \cite{liebetrau1983measures}.

The statistical analysis revealed that the most substantial differences appeared in five key patterns: \textit{Payable}, \textit{Borrower}, \textit{Delegator}, \textit{Implementer}, and \textit{Storage Saver}. \textit{Storage Saver} showed significant differences across all platform pairs ($p < 0.001$), similar to \textit{Borrower}. \textit{Delegator} pattern showed more nuanced distributions, with one non-significant platform pair (Fantom vs. Polygon, $\chi^2 = 2.21$, $p = 0.137$). While our descriptive statistics suggested variations in security patterns like \textit{Reentrancy Guard}, these differences did not meet our effect size threshold.

The variations in \textit{Storage Saver} patterns could reflect different gas cost structures across platforms, while the distinct \textit{Borrower} usage patterns on Arbitrum ($\chi^2 > 1000$, $p < 0.001$) might indicate platform-specific development practices. However, these hypotheses require further investigation.

\begin{tcolorbox}[right=0.1cm,left=0.1cm,top=0.1cm,bottom=0.1cm]
\textit{\textbf{(RQ3) How do micro-pattern frequencies vary across different blockchain platforms?}}
While descriptive statistics suggest variations across many micro-patterns, statistical analysis identifies significant platform-specific differences primarily in 
\textit{Payable},
\textit{Borrower}, \textit{Delegator}, \textit{Implementer}, and \textit{Storage Saver} micro-patterns. This demonstrates our micro-pattern catalog's ability to detect meaningful variations in development practices, though determining the underlying causes requires further research.
\end{tcolorbox}

\subsection{Pattern Relationship Analysis (RQ4)}

The relationships between micro-patterns reveal whether our catalog captures distinct design aspects or overlapping concepts. For each entity $e$ and micro-pattern $mp$, the micro-pattern matching produces a binary outcome $M(mp,e)$ representing micro-pattern presence/absence (see Figure~\ref{fig:definitions}). We analyzed these relationships across our dataset of verified smart contracts through correlation metrics that quantify micro-pattern co-occurrences. 

The relationship between micro-patterns is quantified using a correlation matrix $C$ of size $|P| \times |P|$, where each element $C(mp_i, mp_j)$ represents the Phi coefficient ($\phi$) between micro-patterns $mp_i$ and $mp_j$. We selected $\phi$ as it specifically measures correlation between binary variables:
\[
\phi(mp_i, mp_j) = \frac{n_{11}n_{00} - n_{10}n_{01}}{\sqrt{n_{1\cdot}n_{0\cdot}n_{\cdot1}n_{\cdot0}}}
\]
where $n_{ij}$ represents the count of smart contracts with micro-patterns $i$ and $j$ present (1) or absent (0). Based on established guidelines for binary correlation interpretation \cite{cohen2013statistical}, we considered $\phi > 0.50$ as moderately strong correlation (explaining over 25\% of shared variance) and $\phi < 0.15$ as weak correlation (less than 2\% shared variance). No patterns exhibited strong correlations ($\phi > 0.635$, explaining over 40\% shared variance).

Our analysis revealed predominantly weak correlations between micro-patterns, as reported in Figure \ref{fig:pattern-relationships}. Across \emph{all} platforms, only one pair consistently showed moderately strong correlations ($\phi > 0.50$): \textit{Modifier Usage--Muted} ($\phi \approx 0.54$--$0.60$). Meanwhile, \textit{Operator} --\textit{Supporter} ($\phi \approx 0.50$--$0.54$) attained moderately strong correlations \emph{only} on Ethereum, Polygon, and Optimism.
The vast majority of micro-pattern pairs (between 116
to 122 pairs across different chains) showed weak correlations ($\phi < 0.15$). 

This independence stems from both intentional design decisions and Solidity's language constraints. For example, \textit{Reader}-\textit{Operator} micro-patterns are mutually exclusive by design, since they rely on Solidity's \texttt{view} and \texttt{pure} function types respectively, while \textit{Stoppable} and \textit{Ownable} necessarily include \textit{Modifier Usage}, as they implement their functionality using modifiers. This relationship is unidirectional inclusive, which means even if these micro-patterns require modifiers, the presence of \textit{Modifier Usage} does not imply either \textit{Stoppable} or \textit{Ownable}, as modifiers can be used for other purposes in Solidity smart contracts.

\begin{figure}
    \centering
    \includegraphics[width=\linewidth]{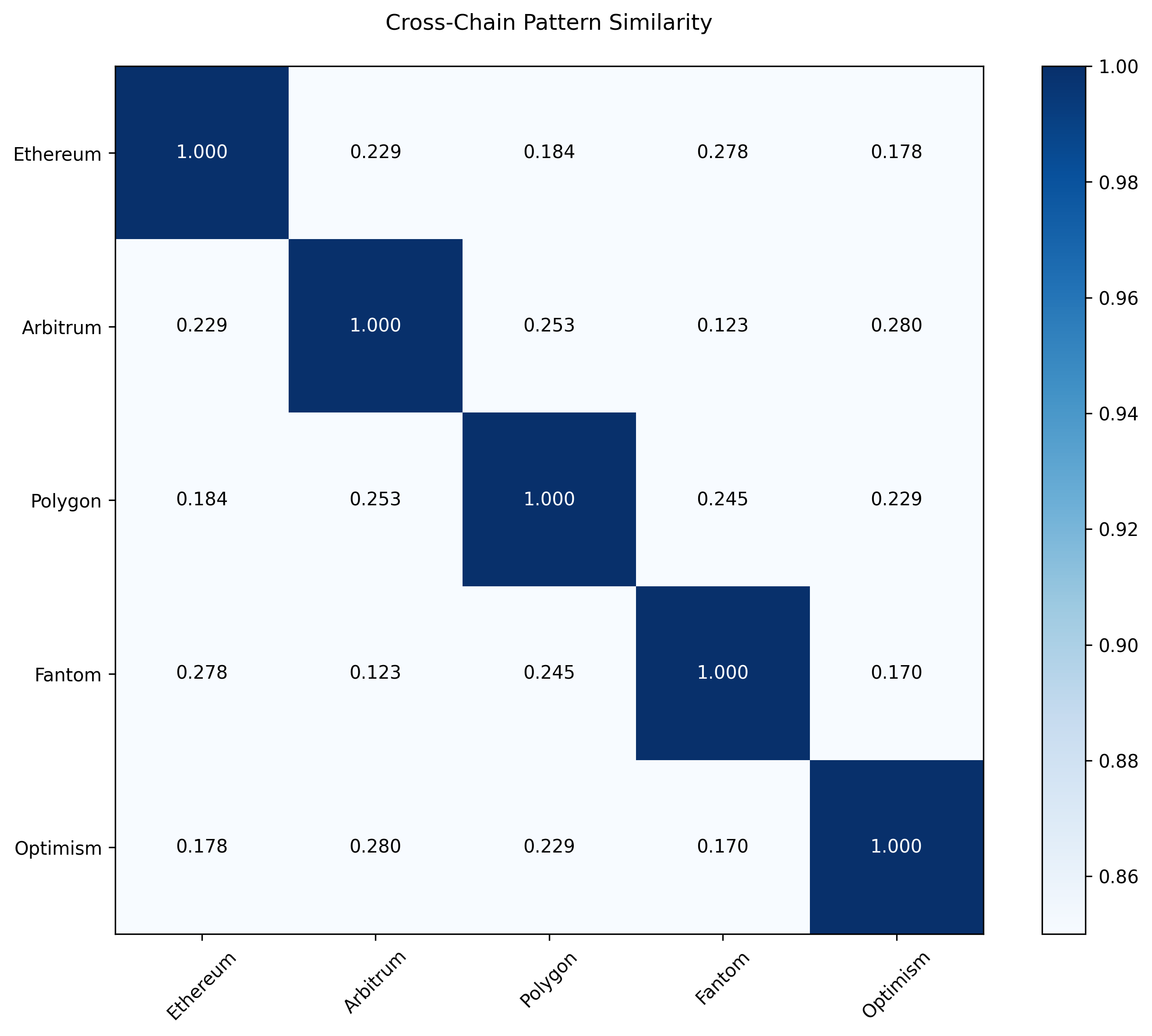}
    \caption{Cross-Chain Pattern similarity}
    \label{fig:spearman-test}
\end{figure}

To assess \emph{relationship stability} across platforms, we used Mantel tests \cite{mantel1967detection} to compare \emph{co-occurrence structures} of micro-patterns on different blockchains. These tests yielded low, statistically nonsignificant correlations ($r = 0.014$--$0.145$, $p > 0.05$), indicating that how patterns co-occur in one blockchain does not strongly align with how they co-occur elsewhere. Additionally, we computed Spearman correlations \cite{spearman1961proof} (see Figure~\ref{fig:spearman-test}) by flattening each chain's pattern-usage profile into a vector and measuring similarity across platforms; those values also remained relatively low ($r \approx 0.18$--$0.28$). Together, these findings support our conclusion that micro-pattern co-occurrences remain effectively independent across different blockchains.

The findings suggested two key insights for smart contract development. First, the widespread weak correlations ($\phi < 0.15$ for $>$115 micro-pattern pairs per platform) indicated that most micro-patterns represent independent design choices that can be flexibly combined according to specific contract requirements. Second, the few moderate correlations we observed ($\phi > 0.50$) likely reflected inherent relationships in contract design, such as when certain functionality naturally requires specific implementation micro-patterns. This independence across our catalog suggested we identified distinct, non-overlapping aspects of smart contract design.

\begin{tcolorbox}[right=0.1cm,left=0.1cm,top=0.1cm,bottom=0.1cm]
\textit{\textbf{(RQ4) Which relationships exist between different micro-patterns, and how are they spread?}}
Our analysis showed that most micro-pattern pairs ($>$115 per platform) were weakly correlated ($\phi < 0.15$), with only a few micro-patterns showing moderate correlations ($\phi > 0.50$). This widespread independence, consistent across all blockchain platforms, suggests our catalog captured distinct, non-overlapping aspects of smart contract design. 
\end{tcolorbox}

\section{Threats to Validity}\label{sec:threats}

\textit{Internal Validity.} Our micro-pattern detection framework achieved a 93\% success rate in processing  25000 verified smart contracts, with failures primarily stemming from dependency resolution and import remapping issues. While successfully processed contracts demonstrate reliable micro-pattern detection, we acknowledge that the dependency-related failures could have influenced our analysis. 

Our focus on publicly verifiable contracts may underrepresent certain patterns, particularly security-focused ones like \textit{Reentrancy Guard}, as security-critical contracts often remain private or use alternative verification platforms. Future work should analyze security-audited contracts from professional sources and audit firms to better understand pattern adoption in high-security contexts.

The framework's core functionality relies on AST parsing and correct dependency resolution, making it sensitive to project structure and compiler versions. To address version compatibility challenges, we implemented a version resolution system that manages multiple Solidity compiler versions. Nevertheless, version-specific language features may affect detection reliability. Initial validation shows promising accuracy across common language features and compiler versions, though future work could expand this validation to cover the full range of Solidity's evolving capabilities.

\textit{External Validity.} Our dataset spans five public blockchains, ranging from 4,217 (Optimism) to 4,900 (Polygon) smart contracts per chain. This selection was based on their transaction volume, providing representation of mainstream development practices. However, our focus on EVM-compatible chains may limit generalizability to other blockchain architectures. Additionally, findings may not generalize to private/permissioned chains which operate under different security and performance constraints.

Our analysis of the most recent contracts per chain means findings may be influenced by contemporary market conditions, trending DeFi protocols, and popular contract templates. This recent snapshot may not fully represent historical development practices or capture the diversity of contract types across platform lifetimes.

\section{Conclusion}
\label{sec:conclusion}
This paper introduced the first catalog of 18 micro-patterns for Solidity code, formally defined across five categories: Security, Functional, Optimization, Interaction, and Feedback. Through our automated detection framework, we analyzed 23258 verified smart contracts deployed across five major blockchain platforms (Ethereum, Polygon, Arbitrum, Fantom, and Optimism). Our analysis revealed that 99\% of contracts \textit{exhibited at least one micro-pattern}, with an average of 2.76 micro-patterns per contract. 

Our findings demonstrated that micro-patterns captured meaningful development practices in Solidity. The \textit{Storage Saver} pattern had the highest adoption with a mean coverage of 84.62\%, while security patterns like \textit{Reentrancy Guard} and \textit{Pull Payment} showed more selective implementation, indicating deliberate selection based on specific contract requirements. The statistical analysis revealed significant platform-specific differences, particularly in \textit{Borrower}, \textit{Implementer}, and \textit{Storage Saver} patterns, suggesting that blockchain characteristics may influence development strategies.

The predominantly weak correlations between micro-patterns ($\phi$ < 0.15 for over 115 micro-pattern pairs) suggested our catalog captured distinct aspects of smart contract design. 
This independence across our catalog indicated we identified non-overlapping design aspects in smart contract implementations.

The integration of micro-pattern analysis into the development process can bridge the gap between traditional software metrics and Solidity's unique requirements. The formal definitions and automated detection framework established in this work will provide a foundation for investigating how micro-patterns influence development practices, supporting future research in automated quality assurance for blockchain systems.

\balance
\bibliographystyle{ACM-Reference-Format}
\newpage
\bibliography{biblio}
\newpage
\begin{figure*}[b!]
\appendix
    \section{Solidity Micro-Patterns Formal Description and Auxiliary Definitions}
\end{figure*}

\begin{table*}[!b]
\label{tab:patterns-logic}
\caption{Solidity Micro-Patterns Formalized in First-Order Logic}
\small
\begin{tabular}{p{2cm}| p{15.5cm}}
\toprule
\rowcolor{lightBg}
\textcolor{headerBlue}{\textbf{Pattern}} & \textcolor{headerBlue}{\textbf{Formal Definition (FOL)}} \\
\midrule
\multicolumn{2}{l}{\textcolor{headerBlue}{\textbf{Security}}} \\
\midrule

\textbf{Ownable} & 
$\exists\,\textcolor{varColor}{v} \in \textcolor{predicateColor}{\mathit{stateVars}}(C),\ 
\exists\,\textcolor{varColor}{m} \in \textcolor{predicateColor}{\mathit{modifiers}}(C),\ 
\exists\,\textcolor{varColor}{f} \in \textcolor{predicateColor}{\mathit{functions}}(C)\colon \bigl(\textcolor{predicateColor}{\mathit{type}}(\textcolor{varColor}{v}) = \textcolor{typeColor}{\texttt{address}}\ \wedge\ \textcolor{predicateColor}{\mathit{isPublic}}(\textcolor{varColor}{v})\bigr)\ \wedge\ \textcolor{predicateColor}{\mathit{checksVar}}(\textcolor{varColor}{m},\textcolor{varColor}{v})\ \wedge\ \textcolor{predicateColor}{\mathit{uses}}(\textcolor{varColor}{f},\textcolor{varColor}{m})$ \\

\textbf{Stoppable} &
$\exists\,\textcolor{varColor}{v} \in \textcolor{predicateColor}{\mathit{stateVars}}(C),\ 
\exists\,\textcolor{varColor}{m} \in \textcolor{predicateColor}{\mathit{modifiers}}(C),\ 
\exists\,\textcolor{varColor}{f} \in \textcolor{predicateColor}{\mathit{functions}}(C)\colon \bigl(\textcolor{predicateColor}{\mathit{type}}(\textcolor{varColor}{v}) = \textcolor{typeColor}{\texttt{bool}}\ \wedge\ \textcolor{predicateColor}{\mathit{isPauseFlag}}(\textcolor{varColor}{v})\bigr)\ \wedge\ \textcolor{predicateColor}{\mathit{checksVar}}(\textcolor{varColor}{m},\textcolor{varColor}{v})\ \wedge\ \textcolor{predicateColor}{\mathit{toggles}}(\textcolor{varColor}{f},\textcolor{varColor}{v})$ \\

\textbf{Pull Payment} &
$\exists,\textcolor{varColor}{m} \in \textcolor{predicateColor}{\mathit{stateVars}}(C),\
\exists,\textcolor{varColor}{f} \in \textcolor{predicateColor}{\mathit{functions}}(C)\colon \bigl(\textcolor{predicateColor}{\mathit{type}}(\textcolor{varColor}{m}) = \textcolor{typeColor}{\texttt{mapping(address}}\Rightarrow\textcolor{typeColor}{\texttt{uint)}}\bigr)\ \wedge\ \textcolor{predicateColor}{\mathit{checksMapping}}(\textcolor{varColor}{m})\ \wedge\ \textcolor{predicateColor}{\mathit{transfersToSender}}(\textcolor{varColor}{f})$ \\

\textbf{Reentrancy Guard} &
$\exists,\textcolor{varColor}{v} \in \textcolor{predicateColor}{\mathit{stateVars}}(C),\
\exists,\textcolor{varColor}{m} \in \textcolor{predicateColor}{\mathit{modifiers}}(C),\ \exists,\textcolor{varColor}{f} \in \textcolor{predicateColor}{\mathit{functions}}(C)\colon \bigl(\textcolor{predicateColor}{\mathit{checksVar}}(\textcolor{varColor}{m},\textcolor{varColor}{v})\ \Rightarrow \textcolor{predicateColor}{\mathit{setVar}}(\textcolor{varColor}{m},\textcolor{varColor}{v})\ \Rightarrow \textcolor{predicateColor}{\mathit{execute}}(\textcolor{varColor}{f})\ \Rightarrow \textcolor{predicateColor}{\mathit{setVar}}(\textcolor{varColor}{m},\textcolor{varColor}{v})\bigr)$ \\

\midrule
\multicolumn{2}{l}{\textcolor{headerBlue}{\textbf{Functional}}} \\
\midrule

\textbf{Payable} &
$\exists\,\textcolor{varColor}{f_i},\textcolor{varColor}{f_j} \in \textcolor{predicateColor}{\mathit{functions}}(C)\colon  \bigl((\textcolor{predicateColor}{\mathit{name}}(\textcolor{varColor}{f_i}) = \textcolor{typeColor}{\texttt{"fallback"}}) \wedge\ (\textcolor{predicateColor}{\mathit{name}}(\textcolor{varColor}{f_j}) = \textcolor{typeColor}{\texttt{"receive"}})\bigr)$ \\

\textbf{Borrower} &
$\exists\,\textcolor{varColor}{l} \in \textcolor{predicateColor}{\mathit{imports}}(C),\ 
\exists\,\textcolor{varColor}{f} \in \textcolor{predicateColor}{\mathit{functions}}(C)\colon \bigl(\textcolor{predicateColor}{\mathit{isLibrary}}(\textcolor{varColor}{l})\ \wedge\ \textcolor{predicateColor}{\mathit{uses}}(\textcolor{varColor}{f},\textcolor{varColor}{l})\bigr)\ $ \\

\textbf{Implementer} &
$\forall,\textcolor{varColor}{f} \in \textcolor{predicateColor}{\mathit{functions}}(C)\colon \exists,\textcolor{varColor}{i} \in \textcolor{predicateColor}{\mathit{inherited}}(C)\colon \bigl(\textcolor{predicateColor}{\mathit{implements}}(\textcolor{varColor}{f},\textcolor{varColor}{i})\bigr)$ \\

\textbf{Modifier Usage} &
$\exists\,\textcolor{varColor}{m} \in \textcolor{predicateColor}{\mathit{modifiers}}(C),\ 
\exists\,\textcolor{varColor}{f} \in \textcolor{predicateColor}{\mathit{functions}}(C)\colon \textcolor{predicateColor}{\mathit{uses}}(\textcolor{varColor}{f},\textcolor{varColor}{m})$ \\

\midrule
\multicolumn{2}{l}{\textcolor{headerBlue}{\textbf{Optimization}}} \\
\midrule

\textbf{Storage Saver} &
$\forall\,\textcolor{varColor}{v} \in \textcolor{predicateColor}{\mathit{stateVars}}(C)\colon \neg\textcolor{predicateColor}{\mathit{wasteSpace}}(\textcolor{varColor}{v})$ \\

\textbf{Reader} &
$\forall\,\textcolor{varColor}{f} \in \textcolor{predicateColor}{\mathit{functions}}(C)\colon \textcolor{predicateColor}{\mathit{isView}}(\textcolor{varColor}{f})$ \\

\textbf{Operator} &
$\forall\,\textcolor{varColor}{f} \in \textcolor{predicateColor}{\mathit{functions}}(C)\colon \textcolor{predicateColor}{\mathit{isPure}}(\textcolor{varColor}{f})$ \\

\midrule
\multicolumn{2}{l}{\textcolor{headerBlue}{\textbf{Interaction}}} \\
\midrule

\textbf{Provider} &
$\forall\,\textcolor{varColor}{f} \in \textcolor{predicateColor}{\mathit{functions}}(C)\colon \textcolor{predicateColor}{\mathit{isExternal}}(\textcolor{varColor}{f})$ \\

\textbf{Supporter} &
$\forall\,\textcolor{varColor}{f} \in \textcolor{predicateColor}{\mathit{functions}}(C)\colon \textcolor{predicateColor}{\mathit{isInternal}}(\textcolor{varColor}{f})$ \\

\textbf{Delegator} &
$\exists\,\textcolor{varColor}{c} \in \textcolor{predicateColor}{\mathit{stateVars}}(C),\ 
\exists\,\textcolor{varColor}{f} \in \textcolor{predicateColor}{\mathit{functions}}(C)\colon \bigl(\textcolor{predicateColor}{\mathit{isContract}}(\textcolor{predicateColor}{\mathit{type}}(\textcolor{varColor}{c}))\ \wedge\ \textcolor{predicateColor}{\mathit{delegatesTo}}(\textcolor{varColor}{f},\textcolor{varColor}{c})\bigr)\ $ \\

\midrule
\multicolumn{2}{l}{\textcolor{headerBlue}{\textbf{Feedback}}} \\
\midrule

\textbf{Named Return} &
$\forall\,\textcolor{varColor}{f} \in \textcolor{predicateColor}{\mathit{functions}}(C)\colon \textcolor{predicateColor}{\mathit{hasNamedReturns}}(\textcolor{varColor}{f})$ \\

\textbf{Returnless} &
$\forall\,\textcolor{varColor}{f} \in \textcolor{predicateColor}{\mathit{functions}}(C)\colon \textcolor{predicateColor}{\mathit{returnCount}}(\textcolor{varColor}{f}) = 0$ \\

\textbf{Emitter} &
$\forall\,\textcolor{varColor}{f} \in \textcolor{predicateColor}{\mathit{functions}}(C)\colon \textcolor{predicateColor}{\mathit{hasEvent}}(\textcolor{varColor}{f})$ \\

\textbf{Muted} &
$\forall\,\textcolor{varColor}{f} \in \textcolor{predicateColor}{\mathit{functions}}(C)\colon \neg\,\textcolor{predicateColor}{\mathit{hasEvent}}(\textcolor{varColor}{f})$ \\
\bottomrule
\end{tabular}
\centering
\end{table*}

\begin{figure*}[!b]
\label{fig:aux-functions-subset}
\noindent
\small
\begin{minipage}{\textwidth}
  \begin{minipage}[t]{0.45\textwidth}
    \begin{tabular}{p{2.2cm} p{5.5cm}}
    \toprule
    \rowcolor{lightBg}
    \multicolumn{2}{l}{\textcolor{headerBlue}{\textbf{Contract Accessors}}} \\
    \midrule
    \textcolor{predicateColor}{$\mathit{stateVars}$}$(C)$ & \textcolor{defnGray}{Returns the set of state variables in $C$.} \\
    \textcolor{predicateColor}{$\mathit{functions}$}$(C)$ & \textcolor{defnGray}{Returns the set of functions in $C$.} \\
    \textcolor{predicateColor}{$\mathit{modifiers}$}$(C)$ & \textcolor{defnGray}{Returns the set of custom modifiers in $C$.} \\
    \textcolor{predicateColor}{$\mathit{enums}$}$(C)$ & \textcolor{defnGray}{Returns the set of enums declared in contract $C$.} \\
    \textcolor{predicateColor}{$\mathit{imports}$}$(C)$ & \textcolor{defnGray}{Returns the set of libraries or external contracts imported by $C$.} \\
    \textcolor{predicateColor}{$\mathit{storageSlots}$}$(C)$ & \textcolor{defnGray}{Returns the set of storage slots used by $C$.} \\
    \bottomrule
    \end{tabular}

    \begin{tabular}{p{2.2cm} p{5.5cm}}
    \toprule
    \rowcolor{lightBg}
    \multicolumn{2}{l}{\textcolor{headerBlue}{\textbf{Checks and Delegation}}} \\
    \midrule
    \textcolor{predicateColor}{$\mathit{toggles}$}$(\textcolor{varColor}{f},\textcolor{varColor}{v})$ & \textcolor{defnGray}{True if $\textcolor{varColor}{f}$ toggles (switches) a bool variable $\textcolor{varColor}{v}$.} \\
    \textcolor{predicateColor}{$\mathit{delegatesTo}$}$(\textcolor{varColor}{f},\textcolor{varColor}{c})$ & \textcolor{defnGray}{True if $\textcolor{varColor}{f}$ delegates calls to contract $\textcolor{varColor}{c}$.} \\
    \textcolor{predicateColor}{$\mathit{checksVar}$}$(\textcolor{varColor}{m},\textcolor{varColor}{v})$ & \textcolor{defnGray}{True if modifier $\textcolor{varColor}{m}$ checks the value of variable $\textcolor{varColor}{v}$.} \\
    \textcolor{predicateColor}{$\mathit{checksMapping}$}$(\textcolor{varColor}{m},\textcolor{varColor}{v})$ & \textcolor{defnGray}{True if modifier $\textcolor{varColor}{m}$ checks mapping variable $\textcolor{varColor}{v}$ for permissions.} \\
    \textcolor{predicateColor}{$\mathit{checksLock}$}$(\textcolor{varColor}{m},\textcolor{varColor}{v})$ & \textcolor{defnGray}{True if modifier $\textcolor{varColor}{m}$ checks a reentrancy lock variable $\textcolor{varColor}{v}$.} \\
    \bottomrule
    \end{tabular}
  \end{minipage}
  \hfill
  \begin{minipage}[t]{0.52\textwidth}
    \begin{tabular}{p{2.5cm} p{6.5cm}}
    \toprule
    \rowcolor{lightBg}
    \multicolumn{2}{l}{\textcolor{headerBlue}{\textbf{Function/Variable Properties}}} \\
    \midrule
    \textcolor{predicateColor}{$\mathit{isPure}$}$(\textcolor{varColor}{f})$ & \textcolor{defnGray}{True if $\textcolor{varColor}{f}$ is declared \textcolor{typeColor}{\texttt{pure}}.} \\
    \textcolor{predicateColor}{$\mathit{hasExternalCall}$}$(\textcolor{varColor}{f})$ & \textcolor{defnGray}{True if $\textcolor{varColor}{f}$ performs an external call.} \\
    \textcolor{predicateColor}{$\mathit{isView}$}$(\textcolor{varColor}{f})$ & \textcolor{defnGray}{True if function $\textcolor{varColor}{f}$ is declared \textcolor{typeColor}{\texttt{view}} (read-only).} \\
    \textcolor{predicateColor}{$\mathit{returnCount}$}$(\textcolor{varColor}{f})$ & \textcolor{defnGray}{Number of return values in $\textcolor{varColor}{f}$.} \\
    \textcolor{predicateColor}{$\mathit{hasNamedReturns}$}$(\textcolor{varColor}{f})$ & \textcolor{defnGray}{$\textcolor{varColor}{f}$ has named returns.} \\
    \textcolor{predicateColor}{$\mathit{overridesAbstract}$}$(\textcolor{varColor}{f})$ & \textcolor{defnGray}{$\textcolor{varColor}{f}$ overrides an abstract method.} \\
    \textcolor{predicateColor}{$\mathit{type}$}$(\textcolor{varColor}{v})$ & \textcolor{defnGray}{Returns the type of variable $\textcolor{varColor}{v}$.} \\
    \bottomrule
    \end{tabular}

    \begin{tabular}{p{2.5cm} p{6.5cm}}
    \toprule
    \rowcolor{lightBg}
    \multicolumn{2}{l}{\textcolor{headerBlue}{\textbf{Boolean Flags/Storage}}} \\
    \midrule
    \textcolor{predicateColor}{$\mathit{wasteSpace}$}$(\textcolor{varColor}{v})$ & \textcolor{defnGray}{True if $\textcolor{varColor}{v}$ can be placed in a previously non-full slot.} \\
    \textcolor{predicateColor}{$\mathit{size}$}$(\textcolor{varColor}{s})$ & \textcolor{defnGray}{Byte size of storage slot $\textcolor{varColor}{s}$.} \\
    \textcolor{predicateColor}{$\mathit{isLibrary}$}$(\textcolor{varColor}{l})$ & \textcolor{defnGray}{True if $\textcolor{varColor}{l}$ is a Solidity library.} \\
    \textcolor{predicateColor}{$\mathit{isPauseFlag}$}$(\textcolor{varColor}{v})$ & \textcolor{defnGray}{True if $\textcolor{varColor}{v}$ is a boolean flag for a paused/stopped state.} \\
    \textcolor{predicateColor}{$\mathit{isLock}$}$(\textcolor{varColor}{v})$ & \textcolor{defnGray}{True if $\textcolor{varColor}{v}$ is used as a reentrancy lock.} \\
    \textcolor{predicateColor}{$\mathit{transfersEther}$}$(\textcolor{varColor}{f})$ & \textcolor{defnGray}{True if function $\textcolor{varColor}{f}$ transfers Ether.} \\
    \textcolor{predicateColor}{$\mathit{hasEvent}$}$(\textcolor{varColor}{f})$ & \textcolor{defnGray}{True if function $\textcolor{varColor}{f}$ emits an event.} \\
    \bottomrule
    \end{tabular}
  \end{minipage}
\end{minipage}
\\[1em]
\textit{\footnotesize A more exhaustive set of definitions is provided in our replication package, including visibility checks, event-related properties, and additional internal predicates.}
\caption{Representative Subset of Auxiliary Functions and Predicates}

\end{figure*}

\end{document}